# gearshifft – The FFT Benchmark Suite for Heterogeneous Platforms

Peter Steinbach

Max Planck Institute of Molecular Cell Biology and Genetics,
01307 Dresden, Germany,
`steinbac@mpi-cbg.de`

Matthias Werner

Center for Information Services and High Performance Computing,
TU Dresden, 01062 Dresden, Germany
`Matthias.Werner1@tu-dresden.de`

July 11, 2017

Fast Fourier Transforms (FFTs) are exploited in a wide variety of fields ranging from computer science to natural sciences and engineering. With the rising data production bandwidths of modern FFT applications, judging best which algorithmic tool to apply, can be vital to any scientific endeavor. As tailored FFT implementations exist for an ever increasing variety of high performance computer hardware, choosing the best performing FFT implementation has strong implications for future hardware purchase decisions, for resources FFTs consume and for possibly decisive financial and time savings ahead of the competition. This paper therefor presents `gearshifft`, which is an open-source and vendor agnostic benchmark suite to process a wide variety of problem sizes and types with state-of-the-art FFT implementations (`fftw`, `clFFT` and `cuFFT`). `gearshifft` provides a reproducible, unbiased and fair comparison on a wide variety of hardware to explore which FFT variant is best for a given problem size.

# 1 Introduction

Fast Fourier transforms (FFTs, [31]) are at the heart of many signal processing and phase space exploration algorithms. Examples for their substantial usage include image reconstruction in life sciences [27, 28], amino acid sequence alignment in bioinformatics [22], phase space reduction for weather simulations [23], option price analysis and prediction in financial mathematics [19] and machine learning [5] to just name a few.

An FFT is a fast implementation of the discrete Fourier transform which is a standard text-book mathematical procedure. The forward transform is a mapping from an array $x$ of $n$ complex numbers in the time domain to an array $X$ of $n$ complex numbers in the frequency domain (referred to as Fourier domain):

$$X[k] = \sum_{j=0}^{n-1} x[j] e^{\frac{-2\pi \mathrm{i} j k}{n}} \qquad (1)$$

with $k$ being an integer index within $0 \leq k < n$ and the imaginary unit $\mathrm{i}^2 = -1$. This operation was found to be computable in $\mathcal{O}(n \log n)$ complexity by Cooley-Turkey [8], who rediscovered findings of Gauss [16]. The basis of the Cooley-Turkey approach is the observation that the DFT of size $n$ can be rewritten by smaller DFTs of size $n_1$ and $n_2$ by the factorization of $n = n_1 n_2$. Given the indices $j = j_1 n_2 + j_2$ and $k = k_1 + k_2 n_1$, Eq. (1) can be re-expressed as:

$$X[k_1 + k_2 n_1] = \sum_{j_2=0}^{n_2-1} \left( \left( \sum_{j_1=0}^{n_1-1} x[j_1 n_2 + j_2] e^{\frac{-2\pi \mathrm{i} j_1 k_1}{n_1}} \right) e^{\frac{-2\pi \mathrm{i} j_2 k_1}{n}} \right) e^{\frac{-2\pi \mathrm{i} j_2 k_2}{n_2}} \qquad (2)$$

Eq. (2) describes a decomposition that can be performed recursively [15]. Here, $n_1$ is denoted *radix* as it refers to $n_1$ transforms of size $n_2$. These smaller transforms are combined by a *butterfly* graph with $n_2$ DFTs of size $n_1$ on the outputs of the corresponding sub-transforms. Radix-2 DFTs ($n$ being a power of two) are mostly implemented with the Cooley-Tukey algorithm [8]. Stockham's formulations of the FFT can be applied [29] to avoid incoherent memory accesses. Arbitrary and mixed radices can be tackled with the prime-factorization or Chirp Z-transform implemented by the Bluestein's algorithm [6].

The top ten list of the fastest worldwide computer installations (Top500 [24]) shows that the used hardware is by far not homogeneous in terms of vendor and composition. This trend can be even more observed in practice, where library architects and domain specialists are confronted with an essential question: Which FFT implementation works best on what hardware?

With increasing experimental data production [18] and simulation output bandwidths [23], input data to FFT libraries in the order of gigabytes becomes the standard. With the advent of graphics processing units (GPUs) for scientific computing around the beginning of the 21st century and the subsequent availability of general purpose programming paradigms to program these [11], vendor-specific and open-source libraries to perform FFTs on accelerators emerged (`cuFFT` [25] by Nvidia, open-source `clFFT` [3]) to offer performance which supersedes traditional high-performance implementations running on standard Central Processing Units (CPUs) such as the open-source `fftw` library [15] or the Intel specific MKL [20].

To our surprise, comprehensive and peer-reviewed benchmarks of FFT implementations across different hardware platforms have not been published extensively. Either only specific hardware is chosen for the benchmark [2, 12, 26] or only specific FFT implementation variants are tested [9, 10]. In addition, many performance benchmarks are tied to domain-specific implementations [14] that either lack comprehensiveness or the ability to map the results obtained to other implementation requirements.

Thus, a new open-source benchmark package called `gearshifft` [17] has been developed. It is able to benchmark available state-of-the-art FFT libraries in a reproducible, automated, comprehensive and vendor-independent fashion on CPUs and GPUs. `gearshifft` helps library authors and domain-specific developers to choose the best FFT library available. The discussion above motivates the following design goals of `gearshifft`:

- open-source and free code

- standardized output format for downstream statistical analysis

- state-of-the-art build system

- open and extensible architecture with generic interface

- community-ready and vendor independent project infrastructure through version control and public accessibility

Given the multitude of mathematical formulations and the heterogeneity of hardware, `gearshifft` approaches the challenge of benchmarking a variety of FFT libraries from a user perspective. This means, that the following parameters should be easy to study:

- FFT dimension and radix-type (e.g. 32×32×32 as radix-2 3D FFT)

- transform kinds, i.e. real-to-complex or complex-to-complex transforms

- precision, i.e. 32-bit or 64-bit IEEE floating point number representation

- memory mode
  - *in-place*: the input data structure is used for storing the output data (low memory footprint and low bandwidth are to be expected)
  - *out-of-place*: where the transformed input is written to a different memory location than where the input resides (high memory footprint and high bandwidth are to be expected)

- transform direction, i.e. forward (from discrete space to frequency space) or backward (from frequency space to discrete space)

2 ImplementationThe remainder of this article is organized as follows: the C++ implementation of `gearshifft` is discussed in Section 2 after an introduction to modern FFT APIs. The largest part of the paper is dedicated to the presentation of first results in Section 3, after which our conclusions are presented in Section 4.

## 2 Implementation

### 2.1 Using a Modern FFT Library

Before discussing the design of `gearshifft`, a brief introduction into the use and application programming interfaces (APIs) of modern FFT libraries is required to illustrate the design choices made. Many FFT libraries today, and particularly those used in this study, base their API on `fftw` 3.0.

Here, in order to execute an FFT on a given pointer to data in memory, a data structure for plans has to be created first using a planner. For this, the FFT problem is defined in terms of rank (1D, 2D or 3D), shape of the input signal (the dimensional extent), type of the input signal (single or double precision of real or complex inputs), type of the transformation (real-to-complex, complex-to-complex, real-to-half-complex) and memory mode of the transformation (in-place versus out-of-place). These parameters describing the FFT problem are then used as input to the planner.

The planner is a piece of code inside `fftw` that tries to find the best suited radix factorization based on the shape of the input signal. By default, it then performs several FFTs derived from the mathematical descriptions discussed in Section 1 on the input data to sample the runtime of different FFT implementations available inside `fftw`. This ensemble of runtimes is then used to find the optimal FFT implementation to use. After the plan has been created, it is used to execute the FFT itself.

Listing 1: Minimal usage example of the `fftw` single precision real-to-complex planner API. Memory management is omitted.

```
1 int shape[] = {32,32,32};
2 fftw_plan r2c_plan = fftw_plan_dft_r2c(
3   /* rank, here 3D     */           3,
4   /* shape of the input */          shape,
5   /* input data array  */           (float *) input_buffer,
6   /* output data array */           (fftwf_complex *) output,
7   /* plan-rigor flag   */           FFTW_ESTIMATE );
8 fftwf_execute(r2c_plan);
```

Listing 1 illustrates the `fftw` API for a single precision real-to-complex out-of-place transform. `fftw` offers the freedom to choose the degree of optimization for finding the most optimal FFT implementation for the signal at hand by means of the planner flag, also referred to as plan rigors. Listing 1 uses the FFTW_ESTIMATE flag as an example, which is described in the `fftw` manual [13]:

> "FFTW_ESTIMATE specifies that, instead of actual measurements of different algorithms, a simple heuristic is used to pick a (probably sub-optimal) plan

*2 Implementation*

Table 1: Methods an FFT client in `gearshifft` has to implement

| constructor | get_alloc_size    | execute_forward |
|-------------|-------------------|-----------------|
| destructor  | get_transfer_size | execute_inverse |
| allocate    | get_plan_size     | upload          |
| destroy     | init_forward      | download        |
|             | init_inverse      |                 |

quickly. With this flag, the input/output arrays are not overwritten during planning."

`fftw` offers five levels for this planning flag, where two further descriptions are given here:

"`FFTW_MEASURE` tells `fftw` to find an optimized plan by actually computing several FFTs and measuring their execution time. Depending on your machine, this can take some time (often a few seconds).
`FFTW_WISDOM_ONLY` is a special planning mode in which the plan is only created if wisdom is available for the given problem, and otherwise a NULL plan is returned."

In `fftw` terminology, *wisdom* is a data structure representing a more or less optimized plan for a given transform. The `fftw_wisdom` binary, that comes with the `fftw` bundle, generates hardware adapted wisdom files, which can be loaded by the wisdom API into any `fftw` application. `cuFFT` and `clFFT` follow this API mostly, only discarding the plan rigors and *wisdom* infrastructure, cp. Listing 2.

Listing 2: Minimal usage example of the `cuFFT` single precision real-to-complex planner API. Memory management is omitted.

```
1 int N = 32;
2 cufftHandle plan;
3 cufftPlan3d(&plan, N, N, N, CUFFT_R2C);
4 cufftExecR2C(plan, input_buffer, output);
```

## 2.2 The Architecture of `gearshifft`

`gearshifft` is developed as an open-source framework using C++ (following the 2014 ISO standard [21]) and the Boost Unit Test Framework (UTF, [7]). One goal is to have a unified benchmark infrastructure and an extensible set of FFT library clients. The benchmark framework is independent of the used FFT library and provides the measuring environment, data handling and processing of results. `gearshifft` involves template meta-programming for a compile-time constant interface between the clients and the benchmark framework. Such a generic approach is necessary to obtain comparable results between FFT libraries and reproducible data for later statistical analysis while keeping code redundancy and overhead at a minimum.

## 2 Implementation

In `gearshifft` a benchmark is meant to collect performance indicators of the operations in Table 1 defining the interface for the FFT clients. Different parameters such as precision, FFT extents, transform variant, device type or FFT library relate to different benchmarks. `gearshifft` controls many of them by command line arguments. The FFT libraries are related to different `gearshifft` binaries (`gearshifft_cufft`, ...). For the full documentation of `gearshifft` the reader is referred to [17].

There are common interfaces for the context management and for the FFT workflow. The user has to implement the context and the FFT client class. The `create` and `destroy` context methods of the client encapsulate time-consuming device and library initialization, which are measured separately and run only once. The library only must be initialized within the FFT client when the library stores plan information (cp. `fftw` wisdoms). The client's context class derives from `ContextDefault` which enables to access and extend the program options.

Listing 3: Required template arguments for FFT client implementation

```
1 template<
2  typename TFFT,      // e.g. gearshifft::FFT_Inplace_Real, ...
3  typename TPrecision, // e.g. double, float, ...
4  size_t   NDim       // 1,..,3
5  /* .. further template types if needed .. */ >
6 struct MyFFTClient;
```

The FFT client implementation in Listing 3 is instantiated once per benchmark run and follows the *resource allocation is initialization* (RAII) idiom [30]. `gearshifft` invokes the FFT client methods listed in Table 1 to perform the benchmarks and to populate the benchmark data. The FFT client can assign user-defined template types to create different FFT client classes to mimic various use cases.

Depending on the FFT library, after a forward transform the same plan handle might be recreated for backward transform. This saves memory as there is only one plan allocated at any point in time. For example, a `cuFFT` plan allocation can be several times bigger than the actual signal data for the FFT. `fftw` can overwrite input and output buffers during the planning phase, when e.g. `FFTW_MEASURE` is used. Afterwards, the buffers can be filled with data. In turn, this plan handle cannot be recreated later on, as the result buffer of the previous plan would be overwritten at plan recreation. `gearshifft`'s compile-time interface supports this use case, where both plans are allocated before the round-trip FFT starts. The `gearshifft` interface also allows library-specific time measurements, which is only implemented for the `cuFFT` library at the moment, where CUDA events measure the runtime on GPU. For `fftw` and `clFFT`, the CPU timer exposed by the C++14 `chrono` header is used.

Listing 4: Define FFT client types for corresponding FFTs

```
1 namespace MyFFT {
2   using Inplace_Real = gearshifft::FFT<
3     gearshifft::FFT_Inplace_Real, MyFFTClient, TimerCPU >;
```

## 2 Implementation

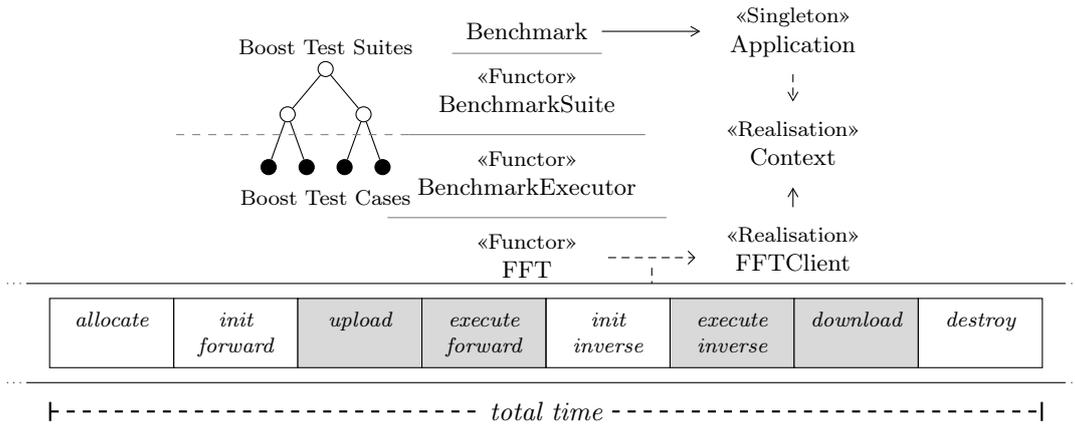

Figure 1: The benchmark framework of `gearshifft` using Boost UTF and a realized FFT interface; Here, only FFT interfaces are shown, that are measured (gray operations are measured by device timers if provided); Context also has an implicit interface, which is omitted here.

Listing 4 shows a type definition for the user implemented class `MyFFTClient` and specifies an in-place-real FFT (cp. Listing 3). This type is added to a list for the benchmark runner, as demonstrated in `benchmark.cpp` (Listing 5). The `gearshifft::List` is a compile-time constant list, which holds the different template instantiations of an FFT client. `FFT_Is_Normalized` denotes a compile time flag if the backward transformed data needs to be normalized in order to achieve identity with the input.

Listing 5: Using FFT client types to run the benchmarks
```
 1 using namespace gearshifft;
 2 using Context           = MyFFT::Context;
 3 using FFTs              = List<MyFFT::Inplace_Real>;
 4 using Precisions        = List<float, double>;
 5 using FFT_Is_Normalized = std::false_type;
 6 int main( int argc, char* argv[] ) {
 7   try {
 8     Benchmark<Context> benchmark;
 9     benchmark.configure(argc, argv);
10     benchmark.run<FFT_Is_Normalized, FFTs, Precisions>();
11   } catch(const std::runtime_error& e) { // ...
```

The back-end of `gearshifft` uses the Boost Unit Test Framework to generate the benchmark instances within a tree data structure, which is referred to as the benchmark tree. The measurement layout and benchmark framework are illustrated in Fig. 1. One single run comprises time measurement of each operation (`allocate`, ...). The total time measures all from `allocate` to `destroy`. The size of the allocated buffers and the memory information of the FFT library (if available) is recorded as well. The functor `FFT` calls the FFT client operations wrapped with time measurements. The input data buffer, filled with a see-saw function in $[0, 1)$ in `BenchmarkData`, is held by the `BenchmarkExecutor`. A copy is given to the `FFT` functor in each run and is used for the output. For each benchmark configuration a number of warmups and benchmark repetitions is performed. After the last benchmark run the round-trip transformed data is validated against the original input data. The error $\varepsilon$ is computed by the sample standard deviation of input and round-trip output. When that error is greater than $10^{-5}$, the benchmark is marked as failed and `gearshifft` continues with the next configuration in the benchmark tree.

`gearshifft` adapts the API of the different FFT libraries to a common interface. The `FFT` functor defines the interface of the common FFT workflow. This pattern refers to Wrapper Facades and Static Adapter design pattern which provides static polymorphism at compile-time [4]. Currently, `gearshifft` implements three different FFT libraries, `cuFFT` (CUDA runtime, [25]) for Nvidia GPUs, `clFFT` (OpenCL runtime, [3]) for CPU and GPUs and `fftw` for CPU (C/C++ runtime, [15]). By this selection, an accelerator-only, a mixed CPU-GPU and a CPU-optimized library is covered. The `cmake` build system is used to setup build paths to construct one executable for each supported FFT library found by `cmake` as well as for collecting the include paths during the build process and library locations for linking later on. There are options for disabling FFT libraries or pointing to non-standard installation paths and to configure compile-time constants such as the error-bound as well as the number of warmups and repetitions.

For the command-line arguments, Boost is utilized, particularly for benchmark list creation and selection. There are several `gearshifft` program options to control benchmark settings, for example:

```
gearshifft_clfft -e 128x128 1024 -r */float/*/Inplace_Real -d cpu
```

Here, the `clFFT` benchmarks would first run a 128×128-point FFT and then a 1024-point FFT, performing in-place transforms with real input data in single-precision. The default setting instructs `gearshifft` to use all CPU cores and to store the results into result.csv. The `gearshifft` benchmark selection syntax supports wildcards. The first wildcard ∗ relates to the title of the FFT client (`ClFFT` in this example). The second one refers to the FFT extents.

## 3 Results

### 3.1 Experimental Environment

This section will discuss the results obtained with `gearshifft v0.2.0` on various hardware in order to showcase the capabilities of `gearshifft`. Based on the applications in [27, 28], 3D real-to-complex FFTs with contiguous single-precision input data are chosen for the experiments. If not stated, this is the transform type assumed for all illustrations hereafter. Expeditions into other use cases will be made where appropriate. The curious reader may rest assured that a more comprehensive study is possible with `gearshifft`, however the mere multiplicity of all possible combinations and use cases of FFT render it neither feasible nor practical to discuss all of them here.

Table 2: Benchmark Hardware

|  | **Taurus** HPC Cluster [33] | **Hypnos** HPC Cluster [1] | **Islay** Workstation | |
|---|---|---|---|---|
| **CPU family** | Haswell Xeon | Sandybridge Xeon | Haswell Xeon | Haswell Xeon |
| **CPU model** | 2× E5-2680 v3 | 2× E5-2450 | 2× E5-2603 v3 | 2× E5-2640 v3 |
| **RAM** | 64 GiB | 48 GiB | 64 GiB | 64 GiB |
| **GPU** (PCIe3.0) | 4x K80 | 2x K20x | 1x P100 | 1x GTX 1080 |
| **GPU memory** | 4x 12 GiB | 6 GiB | 16 GiB | 8 GiB |
| **GPU driver** | 367.48 | 367.48 | 367.48 | 367.57 |
| **OS** | RHEL 6.8 | RHEL 6.8 | Ubuntu 14.04.3 | CentOS 7.2 |

This study concentrates on three modern and current FFT implementations available free of charge: `fftw` (3.3.6pl1, on x86 CPUs), `cuFFT` (8.0.44, on Nvidia GPUs) and `clFFT` (2.12.2, on x86 CPUs or Nvidia GPUs). This is considered as the natural starting point of developers beyond possible domain specific implementations. It should be noted, that this will infer not only a study in terms of hardware performance, but also how well the APIs designed by the authors of `fftw`, `clFFT` and `cuFFT` can be used in practice.

The results presented in the following sections were collected on three hardware installations: All systems presented in Table 2 will be used for the benchmarks in this section. Access was performed via an `ssh` session without running a graphical user interface on the target system. All measurements used the GNU compiler collection (GCC) version 5.3.0 as the underlying compiler. All used GPU implementations on Nvidia hardware interfaced with the proprietary driver and used the infrastructure provided by CUDA 8.0.44 if not stated otherwise. After a warmup step a benchmark is executed ten times. From this, the arithmetic mean and sample standard deviations are used for most of the figures.

## 3.2 Overhead of `gearshifft`

`gearshifft` is designed to be a lightweight framework with a thin wrapper for the FFT clients, where the interface between back-end and front-end is resolved at compile-time. Performance indicators of each benchmark are collected and buffered to be processed after the last benchmark finished. For validation purposes, a `cuFFT` standalone code [17] was created that provides a timer harness like `gearshifft` (referred to as *standalone*). In addition, the time to solution of a straightforward implementation of a round-trip FFT was measured as well (referred to as *standalone-tts*). Both invoke a warm-up step and ten repetitions of the entire round-trip FFT process. Fig. 2 shows the impact of the `gearshifft` internal time measurement with `cuFFT` for two input signal sizes. Fig. 2a illustrates that the time measurement distribution of `gearshifft` overlaps with *standalone* code using multiple timers. A comparison of *gearshifft* and *standalone-tts* visually shows a shift in the average obtained timing result (most likely due to timer object latencies), the scale of this shift resides in the regime below 2 % which we consider negligible. We make this strong claim also because one of the goals of `gearshifft` is measuring individual runs of the benchmark for downstream statistical analysis, thus using one timer object would prohibit this core feature of the benchmark. Fig. 2b shows the impact of larger input signals on the time measurement result. Here, the difference between *gearshifft*, *standalone* and *standalone-tts* decreases even more and converges to a permille level (the longer duration of the benchmark mitigates timer object latencies).

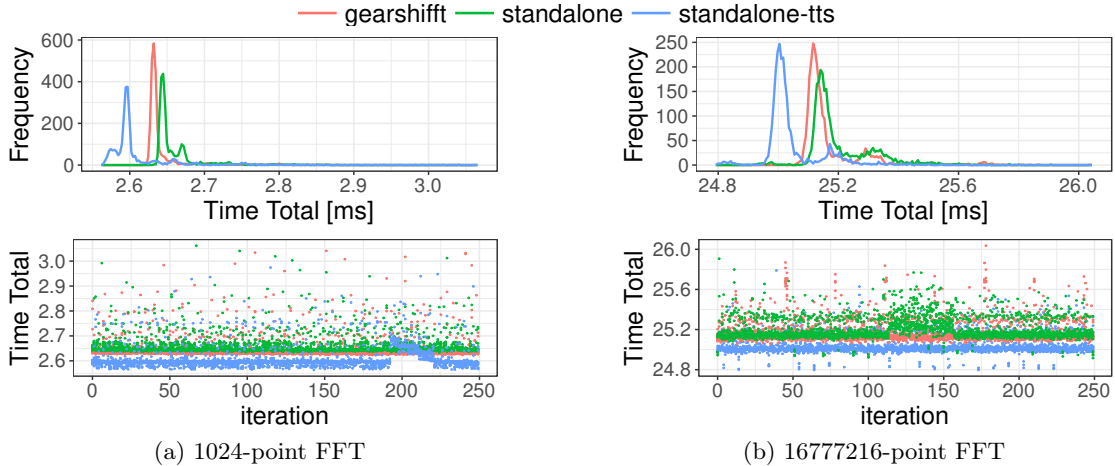

(a) 1024-point FFT

(b) 16777216-point FFT

Figure 2: Time-to-solution measured in *gearshifft* (`cuFFT`), in a *standalone* `cuFFT` application using multiple timer objects and in a standalone application using one timer object (*standalone-tts*) for a single-precision in-place real-to-complex round-trip FFTs on the K80 [33].

## 3.3 Time To Solution

The discussion begins with the classical use case for developers that might be accustomed to small size transforms. As such, an out-of-place transform with `powerof2` 3D signal shapes will be assumed. The memory volume required for this operation amounts to the real input array plus an equally shaped complex output array of the same precision. Fig. 3 reports a comparison of runtime results of `powerof2` single-precision 3D real-to-complex forward transforms from `fftw` and `cuFFT`. It is evident that given the largest device memory available of 16 GiB, the GPU data does not yield any points higher than 8 GiB. The more recent GPU models supersede `fftw` which used all 2×12 CPU Intel Haswell cores. Any judgment on the superiority of `cuFFT` over `fftw` can be considered premature at this point, as `fftw` was used with the `FFTW_ESTIMATE` planner flag.

Fig. 4 compares the time-to-solution to the actual time spent for the FFT operation itself. `FFTW_MEASURE` imposes a total runtime penalty of 1 to 2 orders of magnitude with respect to `FFTW_ESTIMATE`. It however offers superior performance considering FFT execution time compared to `FFTW_ESTIMATE`. To compare `FFTW_ESTIMATE` or `FFTW_MEASURE` with plans using `FFTW_WISDOM_ONLY`, wisdom files are generated with the `fftw_wisdom` binary. `fftw_wisdom` precomputed plans for a canonical set of sizes (powers of two and ten up to $2^{20}$) in `FFTW_PATIENT` mode, which in all took about one day on Taurus [33] using (see [13] for command-line flag details): `fftwf-wisdom -v -c -n -T 24 -o wisdomf`.

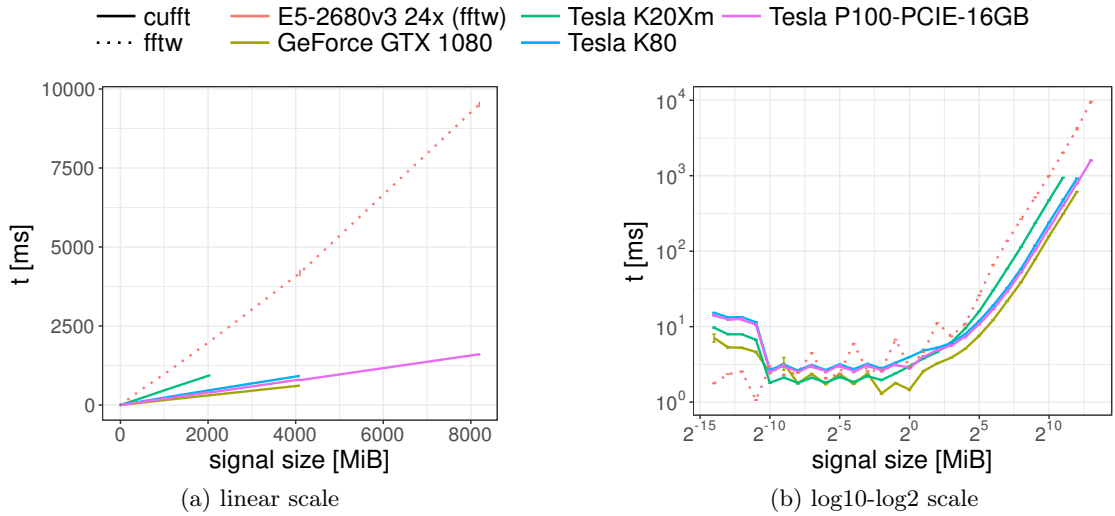

Figure 3: Time-to-solution for `powerof2` 3D single-precision real-to-complex out-of-place forward transforms using `fftw` (`FFTW_ESTIMATE`) and `cuFFT`. Fig. 3b shows the same data as Fig. 3a but in a log10-log2 scale.

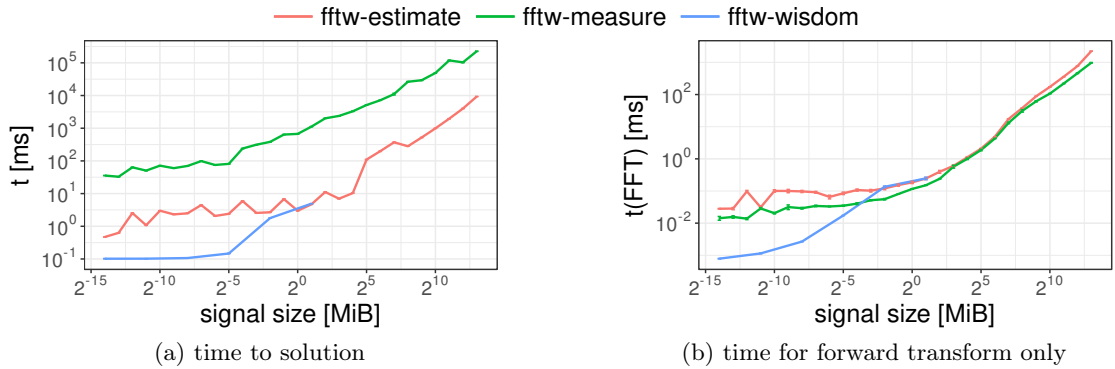

Figure 4: `fftw` on Intel E5-2680v3 CPU with `FFTW_ESTIMATE`, `FFTW_MEASURE` and `FFTW_WISDOM_ONLY` computing `powerof2` 3D single-precision real-to-complex in-place forward transforms. Fig. 4a reports the time to solution, whereas Fig. 4b shows the time spent for the execution of the forward transform only. Both figures use a log10-log2 scale.

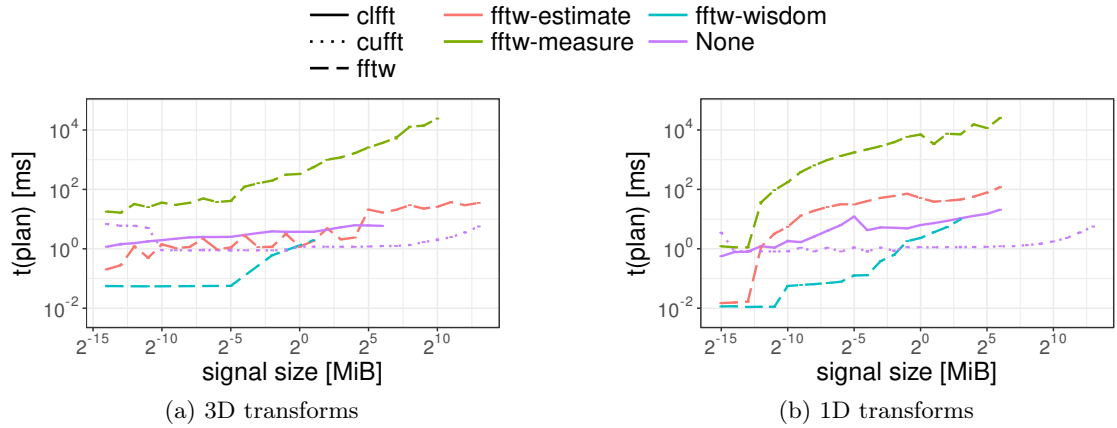

Figure 5: Time-to-plan for `powerof2` single-precision in-place real-to-complex forward transforms using `fftw` (Intel E5-2680v3 CPU), `cuFFT` (K80 GPU) and `clFFT` (K80 GPU). Fig. 5a reports the complete time to plan for 3D FFTs and Fig. 5b for 1D FFTs. "None" refers to the planning with `cuFFT` or `clFFT` as they do not support the plan rigor concept. Both figures use a log10-log2 scale.

As during plan creation, the *wisdom* has to be loaded from disk only, the planning times for calling the planner with `FFTW_WISDOM_ONLY` are drastically reduced. Fig. 4b shows that the user is rewarded by pure FFT runtimes of less than an order of magnitude for small signal sizes. Unexpectedly, the FFT runtimes become larger than those of `FFTW_ESTIMATE` for input signal sizes of more than 32 KiB, which apparently contradicts the `FFTW_PATIENT` setting which should find better plans than `FFTW_MEASURE`. It must be emphasized that the planning times for `FFTW_MEASURE` become prohibitively long and reach minutes for data sets in the gigabyte range. This is a well-known feature of `fftw` as the authors note in [15]:

> "In performance critical applications, many transforms of the same size are typically required, and therefore a large one-time cost is usually acceptable."

`gearshifft` allows one to dissect this problem further and isolate the planning time only. Fig. 5 illustrates the problem to its full extent. `FFTW_MEASURE` consumes up to 3–4 orders of magnitude more planning time than other plan-rigors and plans from GPU based libraries. The 3D planning is compared with its counterpart in 1D (see Fig. 5b). It is important to note that `fftw` planning in 1D appears to be very time consuming as the `FFTW_MEASURE` curve is very steep compared to Fig. 5a. At input sizes of 128 MiB in 1D, the planning phase exceeds the duration of 100 s. The multi-threaded environment could be a problem for `fftw` (compiled against OpenMP): when using 24 threads in `fftw` the time to solution with `FFTW_MEASURE` was up to 6× slower than using 1 thread. Even worse, `FFTW_PATIENT` was up to 50× slower than in a single-thread environment. Unfortunately, the number of threads used for wisdoms, which usually run in `FFTW_PATIENT` mode, must be equal to the ones used by the client later on.

In practice, this imposes a challenge on the client to the `fftw` API. Not only is the time to solution affected by this behavior which is a crucial quantity in FFT-heavy applications. Moreover, in an HPC environment the runtime of applications needs to be known before executing them in order to allow efficient and rapid job placement on compute resources. From another perspective, this asserts a development pressure on the developer interfacing with `fftw` as she has to create infrastructure in order to perform the planning of `fftw` only once and reuse the resulting plan as much as possible. Furthermore, based on these observations of Fig. 4 and Fig. 5 weighing plan time versus execution time, it becomes more and more unclear for a user of `fftw` which plan rigor to use in general.

## 3.4 Comparing CPU versus GPU runtimes

The last section finished by discussing a design artifact, that the `fftw` authors introduced in their API and which other FFT libraries adopted. Another important and common question is whether GPU accelerated FFT implementations are really faster than their CPU equivalents. Although this question cannot be answered comprehensively in our study, there are several aspects to be explored. First of all, modern GPUs are connected via the PCIe bus to the host system in order to transfer data, receive instructions and to be supplied with power. This imposes a severe bottleneck to data transfer and is sometimes neglected during library design. Therefore, the time for data transfer needs to be accounted for or removed from the measurement. `gearshifft`s results data model offers access to each individual step of a transformation, see Fig. 1. Hereby it is possible to isolate the runtime for the FFT transform.

Fig. 6 shows the runtime spent for computing the forward FFT for real single precision input data. This illustration is a direct measure for the quality of the implementation and the hardware underneath. For the 3D case in Fig. 6a `fftw` seems to provide compelling performance if the input data is not larger than 1 MiB on a double socket Haswell Intel Xeon E5 CPU. Above this limit, the GPU implementations offer a clear advantage by up to one order of magnitude. The current Pascal generation GPUs used with `cuFFT` provide the best performance, which does not come by surprise as both cards are equipped with GDDR5X or HBM2 memory which are clearly beneficial for an operation that yields rather low computational complexity such as the FFT. In the 1D case of Fig. 6b, the same observations must be made with even more certainty. The cross-over of `fftw` and the GPU libraries occurs at an earlier point of 64 KiB.

Another observation in Fig. 6a is that the general structure of the runtime curves of GPU FFT implementations follows an inverse roofline curve [32]. That is for input signals smaller than the roofline turning point at 1 MiB the FFT implementation appears to be of constant cost, i.e. to be compute bound. Above the aforementioned threshold, the implementation appears to be memory bound and hence exposes a linear growth with growing input signals which corresponds to the $\mathcal{O}(n \log n)$ complexity observed in Section 1 and validates the algorithmic complexity in [32] as well.

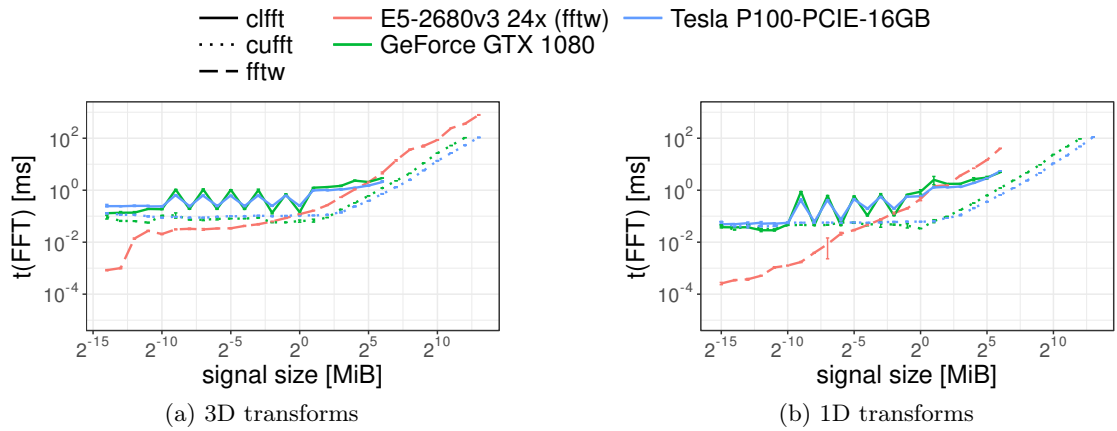

(a) 3D transforms
(b) 1D transforms

Figure 6: Time for computing `powerof2` out-of-place single-precision real-to-complex forward transforms for 3D and for 1D shapes. Both figures use a log10-versus-log2 scale. Curves on the Intel E5-2680v3 based node were obtained with `fftw`, the data on Nvidia GPUs was obtained with `cuFFT` and `clFFT`.

Finally, it is not to our surprise that the `clFFT` results reported in Fig. 6 cannot be considered optimal. As we executed `clFFT` on Nvidia hardware interfacing with the OpenCL runtime coming with CUDA and interfaced to the Nvidia proprietary driver, OpenCL performance can not be considered a first-class citizen in this environment. Only in Fig. 6b, the `clFFT` runtimes are below those of `fftw`. These experiments should be repeated on AMD hardware where the OpenCL performance is expected to be better.

### 3.5 Non-`powerof2` transforms

It is often communicated, that input signals should be padded to `powerof2` shapes in order to achieve the highest possible performance. With `gearshifft` the availability and quality of the common mathematical approaches across many FFT libraries can now be examined in detail. For the sake of brevity, only the results for `fftw` (Intel E5-2680v3 CPU) and `cuFFT` (P100) are presented here.

Fig. 7 confirms that `powerof2` transforms are generally faster than `radix357` and `oddshape` transforms. Excluding the long planning time `fftw` offers the fastest FFT runtime until the turning point at 1 MiB, see Fig. 7a. However, looking at time to solution in Fig. 7b `clFFT` on the CPU outperforms `fftw` by 1 to 2 orders of magnitude due to the long planning times of `fftw`. At very small input signal sizes, `cuFFT` lacks behind `clFFT` on the CPU until 1 KiB for `powerof2` shapes, where `cuFFT` offers superior or comparable runtimes thereafter. `clFFT` only offers support for `powerof2` and `radix357` shape types but has almost the same performance for either. `cuFFT` shows an FFT runtime difference of up to one order of magnitude on the P100 for large input signals (Fig. 7a) of `powerof2` and `oddshape` type, where the time to solution converges due to planning and transfer penalties (Fig. 7a).

For a large range of input signal sizes between $2^{-10}$ MiB to $2^7$ MiB a padding to

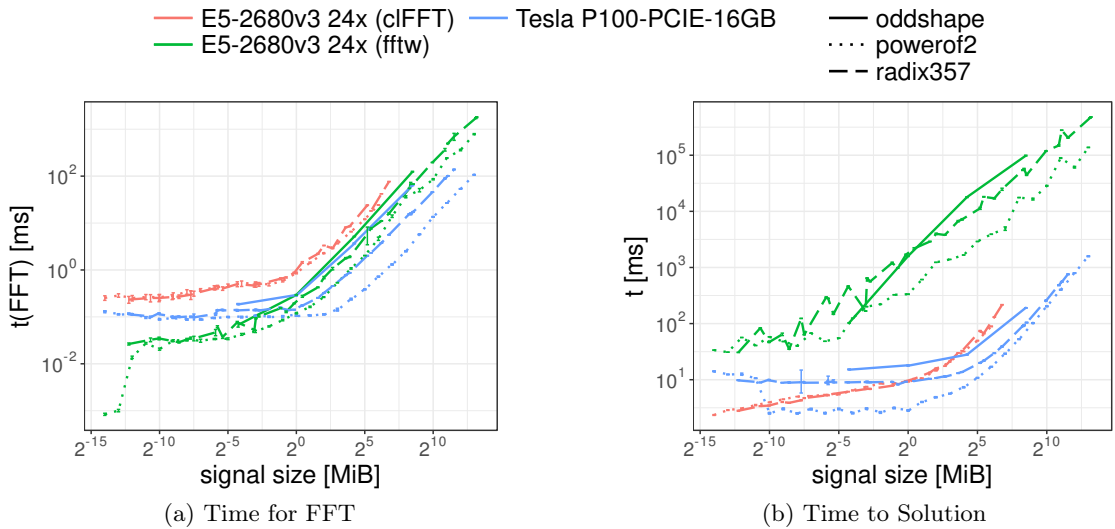

Figure 7: `fftw` and `clFFT` on Intel E5-2680v3 CPU with 24 threads versus `cuFFT` on P100 GPU computing single-precision real-to-complex out-of-place forward transforms of 3D shapes. Both figures use a log10-versus-log2 scale.

`powerof2` might be justified when using `cuFFT` if enough memory is available on the device. For `fftw` non-`powerof2` signals can be padded at signal sizes above $2^{-3}$ MiB = 128 KiB. `clFFT` on CPU is only a good choice, when short planning times are more important than transform runtime. `clFFT` provides similar performance on the P100 as on CPU, but it is not shown here.

### 3.6 Data Types

It is a common practice that complex-to-complex transforms are considered more performant than real-to-complex transforms. Therefore, in order to transform a real input array, a complex array is allocated and the real part of each datum is filled with the signal. The imaginary part of each datum is left at 0.

Fig. 8 restricts itself to larger signal sizes in order to aid the visualization. Note that in Fig. 8a, a data point at the same number of elements of the input signal does have different size in memory. `fftw` exposes a factor of 2 and more of runtime difference for signals larger than $2^{15}$ elements comparing real and complex input data types in Fig. 8a. Below this threshold, the performance can be considered identical except for very small input signals although real FFTs always remain faster than complex ones. The situation is different for `cuFFT`, where the overall difference is smaller in general. In the compute bound region of `cuFFT` (below $2^{19}$ elements), complex transforms perform equally well than real transforms given the observed uncertainties. In the memory bound region (above $2^{19}$ elements), real transforms can be a factor of 2 ahead of complex ones which is clearly related to twice the memory accesses.

If single-precision can be used instead of double-precision, then the possible perfor-mance gain can be estimated by Fig. 8b. On the high grade server GPU, the Nvidia Tesla P100, the performance difference remains around $2\times$ in the memory bound region due to double the memory bandwidth required. The results for `fftw` vary more around 1.5 to 2.5 fold regressions between single and double precision inputs across a wider input signal range.

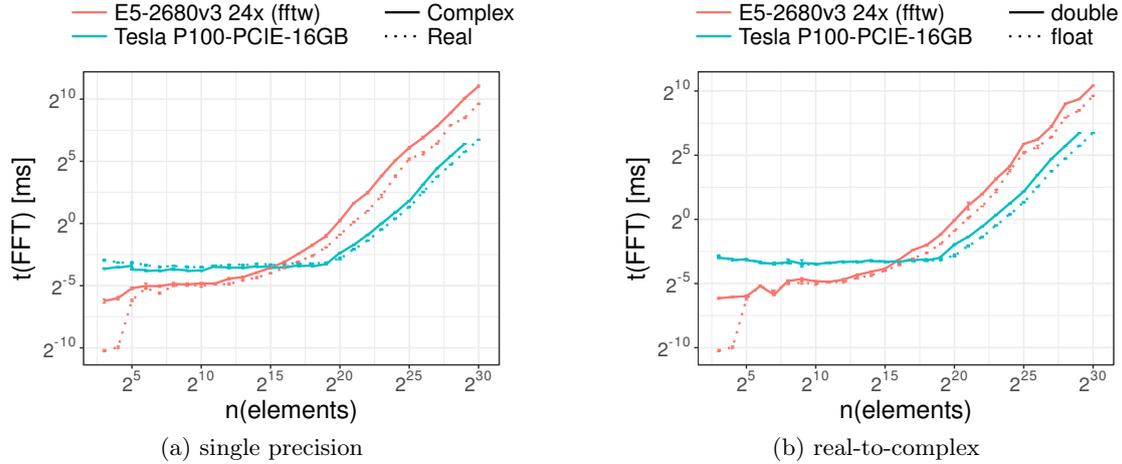

Figure 8: Time for computing a forward FFT using 3D `powerof2` input signals using `fftw` and `cuFFT` on respective hardware versus the number of elements in the input signal. Fig. 8a computes a real-to-complex transform and compares it to a complex-to-complex transform for single precision input data, whereas Fig. 8b shows a real-to-complex transform for either single or double precision. Both figures use a log2-versus-log2 scale.

## 4 Summary

With this paper `gearshifft` is presented to the HPC community and other performance enthusiasts as an open-source, vendor-independent and free FFT benchmark suite for heterogeneous platforms. `gearshifft` is a C++14 modular benchmark code that allows to perform forward and backward FFT transforms on various types of input data (both in shape, memory organization, precision and data type). `gearshifft`'s design offers an extensible architecture to accommodate FFT packages with very low overhead. `gearshifft`'s design choices address both FFT practitioners, FFT library developers, HPC admins or integrators and decision makers supporting a wide range of use cases.

To showcase the capabilities of `gearshifft`, a first study of three common FFT libraries, `fftw`, `clFFT` and `cuFFT` is presented. The performances of CPU based implementations Haswell Xeon CPUs to state-of-the-art Pascal generation Nvidia GPUs are compared. The results indicate that for input signal sizes of less than 1 MiB, the CPU implementation is superior whereas for larger input data size the GPU offers better turn-around. The difference between runtimes of `powerof2`, `radix357` and power-of-19 shaped input data was demonstrated to be negligible for `fftw` and non-negligible for `cuFFT` transforms used in this study. The results further indicate runtime differences when using complex versus real arrays and when comparing double versus single precision data types.

As we warmly welcome contributions of benchmarks from various pieces of hardware, we hope to extend the `gearshifft` repository with many more data sets from platforms used in the HPC arena of today and tomorrow. It is planned to run `gearshifft` on non-x86 hardware to establish a basis for hardware performance comparisons. Connected to this, we plan to explore more state-of-the-art FFT libraries such as Intel IPPS, Intel MKL, AMD's rocFFT, cusFFT etc. It is a future task to consolidate the benchmark data structure and to open another benchmark paths for e.g. FFT callbacks, so that many more analyses are possible than were presented in this paper both in terms of performance exploration as well as energy consumption.

**Acknowledgments.** The work was funded by Nvidia through the GPU Center of Excellence (GCOE) at the Center for Information Services and High Performance Computing (ZIH), TU Dresden, where the K20Xm and K80 GPU cluster Taurus was used. We would like to thank the Helmholtz-Zentrum Dresden-Rossendorf for providing the infrastructure to host the Nvidia Tesla P100 (provided by Nvidia for the GCOE) in the Hypnos HPC cluster. We would also like to thank the Max Planck Institute of Molecular Cell Biology and Genetics for supporting this publication by providing computing infrastructure and service staff working time.